# A complete view of galaxy evolution: panchromatic luminosity functions and the generation of metals

Andrew W Blain (awb@astro.caltech.edu)
Caltech

and

Lee Armus (Spitzer Science Center, Caltech); Frank Bertoldi (Bonn); James Bock (JPL, Caltech); Matt Bradford (JPL); C. Darren Dowell (JPL); Jason Glenn (Colorado); Paul Goldsmith (JPL); Martin Harwit; George Helou (Caltech); J. D. Smith (Toledo); B. T. Soifer (Caltech); Gordon Stacey (Cornell); Joaquin Vieira (Chicago); Min Yun (UMass); Jonas Zmuidzinas (Caltech).

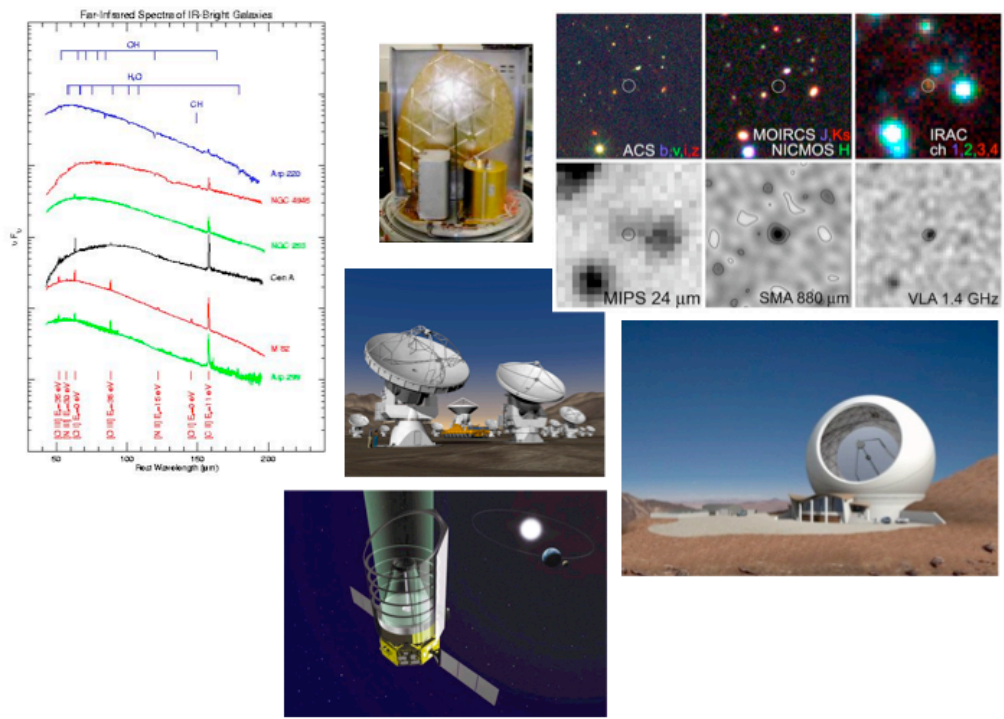

# Abstract

When and how did galaxies form and their metals accumulate? Over the last decade, this has moved from an archeological question to a live investigation: there is now a broad picture of the evolution of galaxies in dark matter halos: their masses, stars, metals and supermassive blackholes. Galaxies have been found and studied in which these formation processes are taking place most vigorously, all the way back in cosmic time to when the intergalactic medium (IGM) was still largely neutral. However, the details of how and why the interstellar medium (ISM) in distant galaxies cools, is processed, recycled and enriched in metals by stars, and fuels active galactic nuclei (AGNs) remain uncertain. In particular, the cooling of gas to fuel star formation, and the chemistry and physics of the most intensely active regions is hidden from view at optical wavelengths, but can be seen and diagnosed at mid- & far-infrared (IR) wavelengths. Rest-frame IR observations are important first to identify the most luminous, interesting and important galaxies, secondly to quantify accurately their total luminosity, and finally to use spectroscopy to trace the conditions in the molecular and atomic gas out of which stars form. In order to map out these processes over the full range of environments and large-scale structures found in the universe - from the densest clusters of galaxies to the emptiest voids – we require tools for deep, large area surveys, of millions of galaxies out to z~5, and for detailed follow-up spectroscopy. The necessary tools can be realized technically. Here, we outline the requirements for gathering the crucial information to build, validate and challenge models of galaxy evolution.

## 1. Introduction: the birth of stars and growth of supermassive blackholes: key remaining questions

The last decade has brought great advances in our phenomenological knowledge about galaxies at redshifts z>1. The GOODS survey has provided a window into the changes in morphologies and the growth in the mass of stars in typical high-redshift galaxies in a small field, while (amongst others) the COSMOS, GEMS and AEGIS surveys have provided shallower, wider views, that start to reach out to encompass the full range of spatial scales in the large-scale structure in the galaxy distribution, which is effectively 5-10 degrees across as projected on the sky at high redshift. Precise observations of the microwave background fluctuations, and large gravitational N-body simulations have determined accurately the nature of the distribution of the (primarily dark) mass in the Universe on 100-Mpc scales. Often targeted initially at optical and UV wavelengths, an array of deep survey fields have attracted the attentions of the most capable facilities at all accessible wavelengths, from the ground and space.

**We have still not answered a key question for understanding the process of galaxy formation: how much total power is radiated by typical galaxies in all environments**?

To determine accurately and reliably the amount of energy that has been released by the millions of galaxies sampled by even existing surveys, it is essential to measure their far-IR radiation. This is dominated in bolometric terms by the thermal emission from dust grains heated to 10-100 K by absorbing optical/UV photons, which the *Infrared Astronomy Satellite* (IRAS), *Infrared Space Observatory (ISO)* and *Spitzer Space Telescope* confirmed produce at least 50% of the total luminosity of galaxies (Dole et al. 2006), and an increasing fraction for the most luminous galaxies. The amount of far-IR emission is difficult or impossible to infer from the spectral properties of the escaping optical/UV light alone: a multi-band/color measurement at restframe far-IR and mm/submm wavelengths is required to determine an accurate measurement of the total power (which peaks at an emitted wavelength of 50-200 μm: see cover spectrum).

The complete far-IR/optical luminosity function of galaxies provides a key test for any theoretical model of galaxy formation. The evolving luminosity function can be integrated and compared across different wavelengths to infer the cumulative build up of stars and metals in galaxies. Restframe far-IR observations are crucial in order to define this function reliably.

**Star formation in galaxies takes place in clouds of molecular gas, and key tracers appear at far-IR wavelengths. Where are the most interesting galaxies, and what processes are taking place within them, in both the spectral and spatial domains?**

The processes responsible for the generation of energy in distant galaxies must of course be probed across a wide range of wavelengths, from radio to hard X-rays. Amongst these, the restframe mid-/far-IR/submm waveband is critical, but currently less accurately investigated. The origin and cycling of metals in the ISM revealed via far-IR observations is crucial for understanding the astrophysical processes that drive galaxy evolution: the dust mass and total luminosity are revealed by continuum observations, and spectral emission from excited molecular and atomic gas reveals the fuel for both star formation, and the conditions around the regions when accretion processes gas onto AGNs (see White Paper submitted by Lee Armus et al.). Some of the most distant luminous AGN have abundances of metals in their cores that already exceed that in the Sun, confirming the significant enrichment of the ISM, at least in the densest and most extreme high-redshift locations. These metals in the ISM, in the form of both dust and gas, are responsible for the opacity of galaxies. By cooling gas in star-forming regions, metals control the formation of all but the very first generation of stars.

The modest optical depth of galaxies at IR wavelengths ensures that even the most heavily embedded regions where star formation and AGN activity takes place are revealed. Furthermore, the IR emission from molecular and atomic gas highlights the star-forming material in the active ISM directly, and is relatively immune to the effects of winds and outflows from the galaxy that are traced using near-IR/optical spectroscopy

Existing observations of far-IR high-redshift galaxies are limited to only the most luminous examples. Moreover, the number of galaxies known in the most luminous classes, those easiest to follow up, is only in the tens. Redshifts and stellar masses for these galaxies can generally be determined from optical and near-IR observations (see Tacconi et al. 2008), but total luminosities, gas masses, and dynamical masses must be derived from far-IR/submm observations of gas and dust. Without this IR information, far-IR dominated high-redshift galaxies appear to be little different from the substantial zoo of optically-selected galaxies at the same redshifts (see Reddy et al. 2008), despite their significantly greater luminosity. Multiwavelength observations remain necessary to infer the properties of high-redshift galaxies: the total power of galaxies in the integrated background light is shared between optical and far-IR photons, but the rarer, most luminous galaxies are most prominent in the far-IR. Direct observations of far-IR emission are thus especially important, if we are to understand the processes that lead to dramatic increases in luminosity for only a modest fraction of the lifetime of the galaxy. It is very likely that we must look to the environment of the galaxy to investigate the triggers for these violent episodes, by identifying far-IR luminous galaxies through the full range of densities in the distribution of galaxies from the richest clusters to the emptiest voids.

2.  **Current knowledge: a reliable, but teasing view of the far-IR Universe**

In the last decade, the legacy of *IRAS & ISO*, and the bulk of the cryogenic phase of the *Spitzer* mission have provided insight into the far-IR Universe at redshifts out to z~1, and substantial information beyond.

Operating with modest-format detectors, and telescopes with apertures only thousands of wavelengths across, submm and far-IR observatories have clear potential for ongoing development: the capability of tools at these wavelengths is growing tremendously, but is still nowhere near mature. The combination of detector area, effective collecting area and improvements in spatial and spectral resolution will reveal in much more detail the process of star formation across all scales from the solar neighborhood right out to the first stars in the universe. Amongst their science capabilities, existing facilities are addressing different aspects of the evolution of the metals in galaxies. These facilities include, on the ground, 10-m class (sub)mm-wave single antenna facilities – the Atacama Pathfinder Experiment (APEX), Atacama Submillimeter Telescope Experiment (ASTE), Caltech Submillimeter Observatory (CSO) and James Clerk Maxwell Telescope (JCMT) - are observing, along with three interferometers: IRAM, SMA and CARMA. The new South Pole Telescope (SPT) and Atacama Cosmology Telescope (ACT) are also operating. The Atacama Large (Sub)Milleter Array (ALMA) is under construction in Chile, the longer-wavelength Large Millimeter Telescope (LMT/GMT) is being built in Mexico, and the wide-field Cornell-Caltech Atacama Telescope (CCAT) is under detailed design study. The Stratospheric Observatory for Infrared Astronomy (SOFIA) is in flight test. The space-borne facilities, *Herschel Space Observatory*, *Planck Surveyor* and *Wide-Field Infrared Space Explorer (WISE)* are all set to fly in 2009, while the *James Webb Space Telescope (JWST)* and Japanese 4-m-class cryogenic-mirror far-IR *SPICA* mission are under construction and study respectively. In the more distant future, larger space-borne apertures such as *CALISTO/SAFIR* and the far-IR interferometers *SPIRIT* and *SPECS* are being studied. The current state of knowledge is illustrated by the following figures.

Fig. 1 illustrates the brightness distribution of galaxies selected at mm/submm wavelengths in a variety of different surveys. The shape of the curves highlights several key points. 1) At bright flux densities, the count is dominated by the far-IR/submm spectrum of ordinary low-redshift galaxies. 2) near 20 mJy, there is a transition to a steeper curve due to a flood of high-redshift galaxies, that appear owing to the unique favorable effect of redshifting the far-IR spectrum into the observing band. The shape of this part of the curve is controlled in large part by the luminosity function of distant far-IR luminous objects. Furthermore, at the foot of this steep rise the fraction of galaxies that are gravitationally lensed could exceed a remarkable 10%. This can provide a new set of laboratories for studying the distribution of dark matter, and a natural boost to telescope sensitivity and angular resolution for follow-up observations. Finally 3) the counts begin to converge at the flux of a typical high-redshift galaxy at ~0.2mJy. This is too faint for existing facilities to probe; however, it is still significantly greater than the flux density expected from a typical optically-selected high-redshift galaxy.

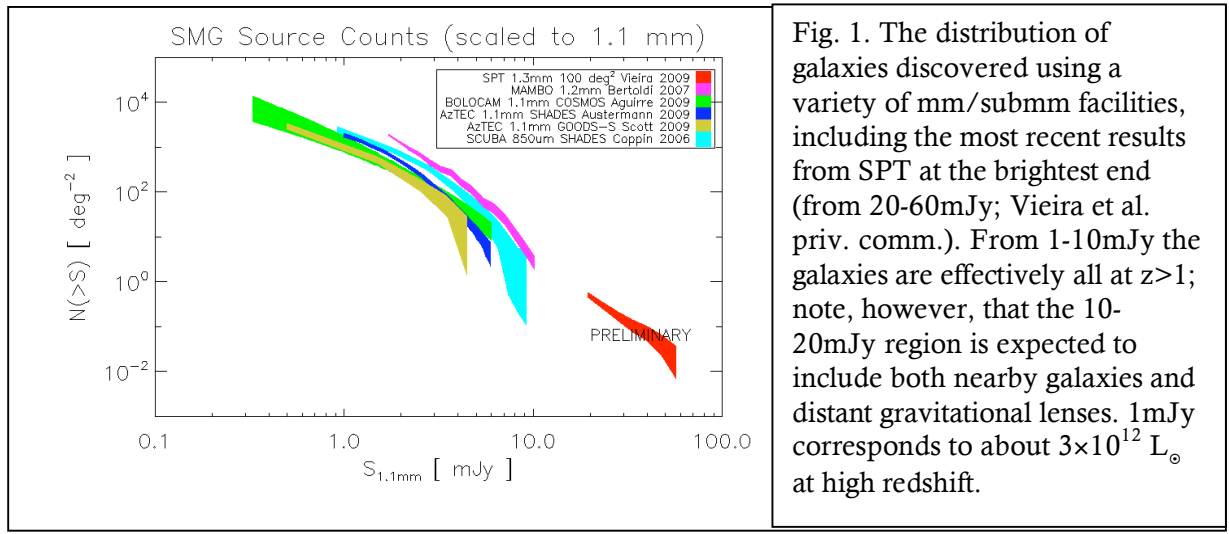

Fig. 1. The distribution of galaxies discovered using a variety of mm/submm facilities, including the most recent results from SPT at the brightest end (from 20-60mJy; Vieira et al. priv. comm.). From 1-10mJy the galaxies are effectively all at z>1; note, however, that the 10-20mJy region is expected to include both nearby galaxies and distant gravitational lenses. 1mJy corresponds to about $3\times10^{12}$ $L_\odot$ at high redshift.

The numbers of galaxies known at the bright and faint end of the curves in Fig. 1 is in the handfuls and hundreds respectively. In order to define the underlying functions precisely, coverage of several bands at submm & far-IR wavelengths is required: we must sample 10's of square degrees at <1mJy and a large fraction of the sky at the brightest end (>100mJy). This will yield samples of $10^{5-6}$ galaxies from 10-deg$^2$ CCAT and 100-deg$^2$ *Herschel* surveys, and several $10^3$ bright distant galaxies from *Planck Surveyor* over the whole sky. The submm selection technique remains effective out to z>10, and so a wide-field survey can yield a sample of very distant targets with red far-IR colors. Amongst these will be the most luminous galaxies formed when the IGM was still neutral: the first cities of stars in the Universe, which probably make up about 1% of the total detections.

After a galaxy is detected, even low-resolution ($R$~100) spectroscopy can be used to determine a redshift, in another waveband, or in the mm/submm/far-IR directly from powerful far-IR atomic/ionic fine structure line emission, from the evenly-spaced ladder of CO rotational emission states, or even perhaps from radio recombination lines. Fig. 2 shows a moderate resolution spectrum made using a prototype wide-band mm-wave spectrograph ZSpec. The target is the Cloverleaf, a well-know, very bright, gravitationally-lensed high-redshift QSO. The relative strength of several different CO rotational lines allows the density and temperature of the molecular gas in the host galaxy to be determined, along with a very precise redshift. Direct spectroscopy at far-IR wavelengths can provide redshifts and diagnostic information for even the most distant galaxies.

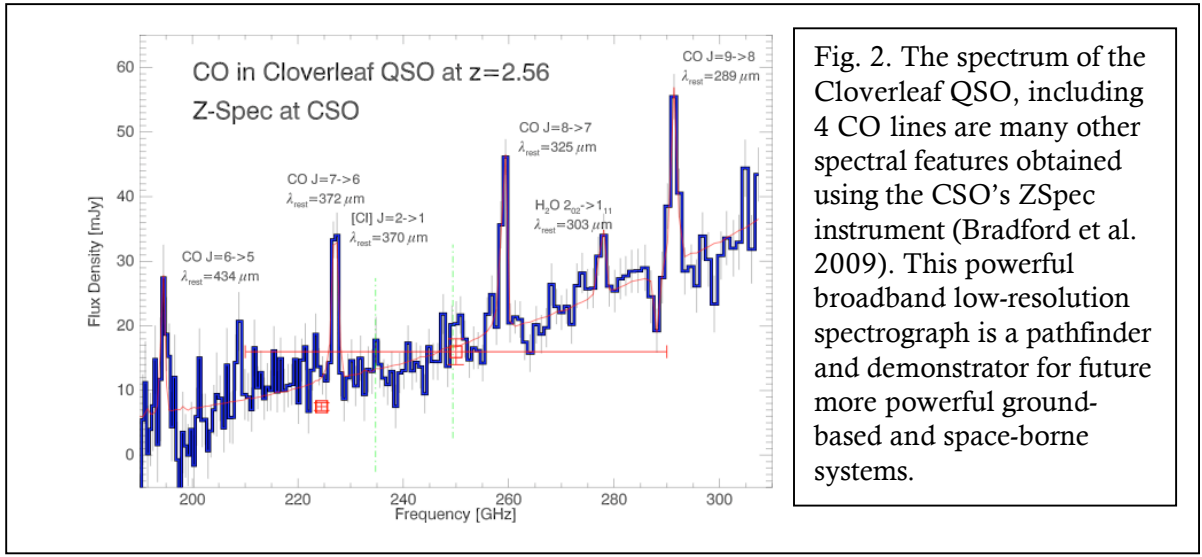

Fig. 2. The spectrum of the Cloverleaf QSO, including 4 CO lines are many other spectral features obtained using the CSO's ZSpec instrument (Bradford et al. 2009). This powerful broadband low-resolution spectrograph is a pathfinder and demonstrator for future more powerful ground-based and space-borne systems.

Detailed interferometric spectral imaging using ALMA can always be used to reveal the mass and excitation conditions of the star-forming regions within. Fig. 3 illustrates the best current view of a specific, very luminous, high-redshift galaxy, exploiting the 0.2-arcsec resolution of the IRAM interferometer to image the continuum and CO emission. The correlations between line velocity and spatial position provide a total mass via dynamics, along with a measurement of the spatial extent of the observed molecular gas. ALMA will be able to make images with a resolution several times finer, in less than an hour.

3. **Requirements to address key questions: finding a wide range of interesting galaxies, and providing the tools to understand their astrophysics**

New facilities coming online will enhance the current observational capabilities illustrated in Figs 2 & 3; however, in order to answer the key science questions posed, additional capabilities are still required. Continuum surveys at (sub)mm/far-IR wavelengths to discover galaxies, and measure their brightness and color must probe deep enough to detect a typical star-forming galaxy, that could be discovered in an optical surveys (with a total luminosity of about $5\times10^{11}L_\odot$), out at least to redshift z~3 where galaxy formation was well underway. ALMA has this capability, but it must search several empty beam-sized fields in order to find each galaxy. Samples of at least $10^5$ galaxies are required in order to mine catalogs to understand the full spread of the properties of the galaxies, to find examples that lie both in the richest clusters and the emptiest voids, and to allow ready cross comparison with the deepest multiwavelength surveys. Even with a very fast instrument, sub-five-arcsec resolution is necessary in a survey to avoid a limit to the depth of observations from 'confusion noise' imposed by the overlapping signals from unresolved galaxies, at sub-mJy levels, and galaxy surface densities of $10^5$ deg$^{-2}$. This demands a ratio of aperture and wavelength $D/\lambda > 4\times10^4$, corresponding to a >15-m antenna in the shortest 350-μm atmospheric window. 3-arcsec resolution is a reasonable requirement to start to determine sizes of the largest galaxies found to date, requiring a 25-m-class antenna (or interferometer baseline). This resolving capability must be coupled with an efficient antenna, and a field of view and detector that can cover 10 deg$^2$ in a reasonable time. The planned 32000-pixel 350-μm SWCam for CCAT could survey to a depth of 0.2mJy RMS over a 10-deg$^2$ field in a practical 2000hr.

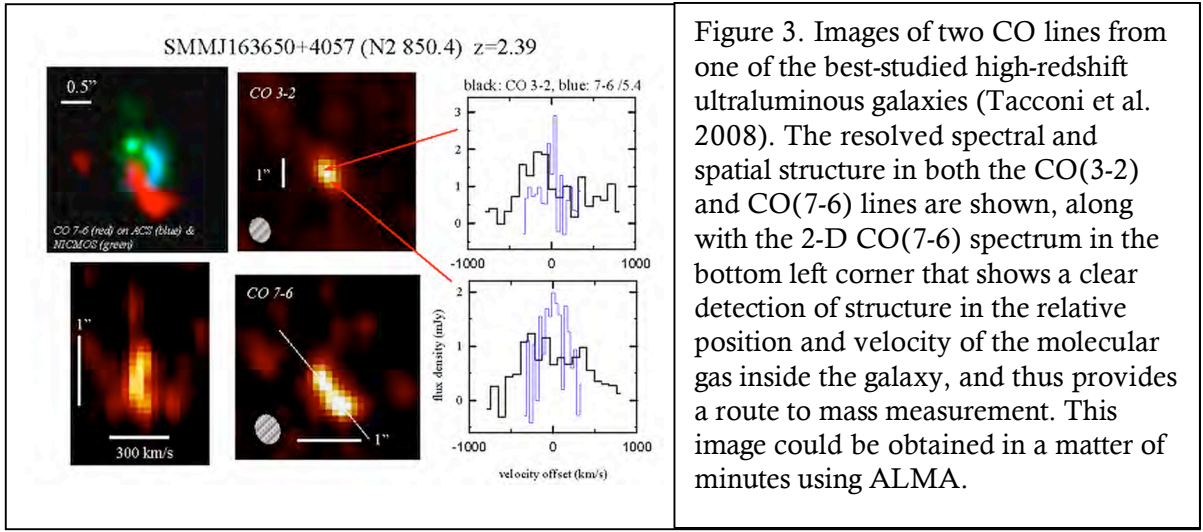

Figure 3. Images of two CO lines from one of the best-studied high-redshift ultraluminous galaxies (Tacconi et al. 2008). The resolved spectral and spatial structure in both the CO(3-2) and CO(7-6) lines are shown, along with the 2-D CO(7-6) spectrum in the bottom left corner that shows a clear detection of structure in the relative position and velocity of the molecular gas inside the galaxy, and thus provides a route to mass measurement. This image could be obtained in a matter of minutes using ALMA.

Multi-color mm-to-far-IR survey observations are crucial, both to provide a red/blue cut that can indicate a coarse redshift, and moreover to provide an accurate measure of luminosity if a redshift is known. A determination of the brightness of a galaxy bluewards of the peak of its spectral energy distribution is thus important: this *requires* access to the submm windows at 350/450 μm from the ground and/or to the far-IR from space.

Spectroscopic probes of the astrophysics in distant galaxies depend on a detection capability at the wavelengths to which specific lines are redshifted, and suitable resolution and sensitivity to enable the observations to be made in a reasonable time. These capabilities are growing rapidly at present, with increasing detector pixel counts, especially with emerging MKID detectors (Glenn et al. 2008), and different dispersive elements being used to demonstrate broadband pseudo-optical-type and radio spectroscopy at mm and submm wavelengths: examples include the ZSPEC, ZSpectrometer, ZEUS and Redshift Search Receiver instruments (see Baker et al. 2007), for which CCAT and LMT are excellent platforms. The development of multi-object mm-wave spectrographs would aid the follow-up investigation of galaxies identified in multi-band mm/submm surveys.

In space, dispersing the spectrum of faint distant galaxies imposes severe but attainable demands on the thermal background from instrument and telescope. Cryogenically-cooled space telescopes are capable of spectroscopy that matches the sensitivity of ALMA, all the way through the far-IR waveband, covering the full range of atomic fine-structure lines (from C, N & O) that extract most energy from the clouds in which stars form. This wide wavelength coverage is very valuable as different energy levels in different species of atomic gas sample a wide range of ionization potentials and critical densities for excitation. This leads to a rich suite of lines that are available to probe the conditions in photodissociation regions around molecular clouds, in X-ray-illuminated dissociation regions, in coronal gas and in the hard radiation fields of AGNs. These fine-structure lines are both barely studied at present and accessible from the ground only in specific atmospheric windows from the very best observing sites. However, the necessary ground-based instruments are real and are demonstrating promising performance (Fig. 2). The potential for learning new things about the physical conditions in distant galaxies using these tools is enormous.

4. **Recommendation and summary**

In order to understand the way in which galaxies in the Universe formed, it is necessary to track the mutually-entangled evolution of star-forming galaxies and AGN, over a wide range of wavelengths. The mm-far-IR wavelength range includes unique diagnostics of both the true luminosities of these galaxies, and the astrophysical conditions within the very parts of galaxies where stars form, by detecting the direct signatures of the gas that turns into stars.

Deep multi-color submm/far-IR images over the full range of spatial scales found in the galaxy distribution will provide a true measure of the star formation rate over the history of the Universe, and highlight how the process depends on environmental conditions. CCAT is a crucial tool for making these measurements that will provide a critical test for all models and theories of the messy details of global star formation.

The necessary resolution, sensitivity, diagnostic capability for the spectral properties of galaxies, and mapping speed can only be obtained using a 25-m-class telescope with new large-format array detectors operating in the highest-frequency atmospheric windows.

A 5-10-m-class cooled telescope with access to the full far-IR spectrum in space can exploit spectroscopic information to moderate the angular resolution requirement.

In order to probe the evolution of galaxies and AGN out to beyond reionization, and look into the regions of the Universe where stars and heavy elements originate, it is essential to discover and study galaxies at restframe far-IR wavelengths.